\documentclass[reprint,amsmath,amssymb,aps]{revtex4-2}

\usepackage{graphicx}
\usepackage{dcolumn}
\usepackage{bm}
\usepackage{braket}

\begin{document}

\title{\bf The thermodynamic stability and phase structure
	of the Einstein-Euler-Heisenberg-AdS black holes}

\author{Yinan Zhao}
\author{Hongbo Cheng}
\email{hbcheng@ecust.edu.cn}
\affiliation{
Department of Physics, East China University of Science and Technology, Shanghai 200237, China\\
The Shanghia Key Laboratory of Astrophysics, Shanghai 200234, China}

\date{\today}

\begin{abstract}
In both canonical ensemble and grand canonical ensemble, the
thermodynamic stability and phase structure of
Einstein-Euler-Heisenberg-AdS black hole are studied. We derive
the Hawking temperature, Helmholtz free energy, Gibbs potential,
entropy and heat capacity of the black holes. We compute the
minimum temperature to find that the phase transition may happen
at the lowest point. The entropy-temperature diagram consists of
two parts. The upper part belonging to the large black holes under
the influence from the electromagnetic self-interactions keeps the
positive heat capacity, leading the huge compact objects to
survive. The lower curves corresponding to the small ones show
that the heat capacity of the tiny black holes is negative, which
means that the nonlinear-effect-corrected smaller sources will
evaporate. The further discussions show that the nonlinear effect
modifies the thermodynamic quantities, but the corrections limited
by the nonlinear factor $\mu$ with allowed values can not change
the properties and the phase structure fundamentally and
thoroughly. We argue that the influence from self-interaction can
not make the Einstein-Euler-Heisenberg-AdS black holes to split
under the second law of thermodynamics.
\end{abstract}

\maketitle


\section{Introduction}

As solutions to the Einstein equations theoretically and being
accepted as compact objects for testing general relativity, the
black holes have received significant attentions for decades
[1-5]. The Event Horizon Telescope (EHT) collaboration proceeded
with the detections of the candidates of black holes such as
supermassive ones at the centers of the galaxy M$87^{\ast}$ and
the Milky Way [6-13]. The black holes have their own
characteristics relating to their structures and environments [1].
The black holes in the background with negative cosmological
constant can be thought as a gauge-gravity duality in the boundary
of Anti de Sitter (AdS) spacetime [14].

According to the astronomical observations, the compact objects
immersed in the magnetic fields produce pair conversion and vacuum
polarization [15, 16], so the influence from the electromagnetic
fields can not be neglected. When the strength of electromagnetic
fields is strong enough, the vacuum seems to act as a material
medium due to the electromagnetic self-interactions for the vacuum
polarization like the light-light scattering or electron-positron
creation [17-19]. The self-interactions can also rewrite the
metric of the compact celestial bodies and leave an imprint in the
background structure around the black holes. More achievements
have been made on the nonlinear electromagnetic effects in the
experiments [20-23]. Within the frame of Euler-Heisenberg theory
with vacuum polarization to one loop, the system Lagrangian with
the Lorentz and gauge invariants is expanded up to the
second-order in the fine structure constant [24, 25]. The
Euler-Heisenberg theory coupling with gravitation leads a static
and spherically symmetric solution for a black hole with electric
or magnetic charge [26-29]. It is interesting to wonder how the
effects from the Euler-Heisenberg issue affect the structure of
spacetime around the black hole. As the possible phenomenon of
birefringence, the two light trajectories were studied in the
background of Einstein-Euler-Heisenberg black hole and the null
geodesics governed by the two optical metrics relating to the
light polarization respectively [30]. The authors of Ref.[30] also
researched on the quasinormal modes of this kind of black holes
characterized by a strong field nearby. It is significant to
generalize the discussion on the Einstein-Euler-Heisenberg-AdS
black holes.

The black hole thermodynamics has attracted a lot of attentions of
physical community for more than twenty years [1]. There has also
been much interest in the investigations of black hole in the AdS
spacetime [1]. The occurrence of phase transition between two
thermodynamically stable black holes and the thermal radiation
from black holes in the AdS world were put forward, which is known
as the Hawking-Page phase transition [31]. The existence of phase
transition between the thermodynamically stable large black hole
and the thermal radiation in the AdS universe was shown nearly
forty years ago [31]. The Hawking-Page phase transition can also
be thought as a confinement/deconfinement process of gauge field
under the AdS/CFT correspondence [14]. It is inevitable that the
black hole thermodynamics may be explored in a new direction with
negative cosmological constant thought as a thermodynamic variable
while the thermodynamic potential enthalpy considered [32-34]. In
this process, the first law of thermodynamics for the
Schwarzschild and Reissner-Nordstrom black holes in an anti-de
Sitter universe originated, and the cosmological constant acts as
a vacuum pressure $P=-\frac{\Lambda}{8\pi}$ conjugating to a
thermodynamic volume $V=\frac{4}{3}\pi r_{+}^{3}$ with horizon
radius $r_{+}$ [32-39]. The efforts have been made that the phase
structures subject to the thermodynamics of Reissner-Nordstrom-AdS
black holes were discussed [40, 41]. The Hawking-Page-like phase
transitions between the AdS black holes were explored [42-55]. The
extension was performed to the P-V phase transitions of the same
kinds of ones [36, 39, 56-62]. In order to gain insights into the
special topic of black hole thermodynamics, we aim to analyze the
thermodynamic stability and phase structure of AdS black holes
[40, 41, 63-66]. It is fundamental to derive and calculate the
basic quantities such as the Hawking temperature and entropy [40,
41]. Further the free energy and the heat capacity of the black
hole in the background with negative cosmological constant are
discussed in the canonical ensemble with fixed charge and the
grand canonical ensemble with fixed potential respectively [40,
41]. The dependence of the free energy on the temperature show the
AdS black hole' phase structure [40, 41]. The nature of the heat
capacity decides the stability of the AdS black hole [40, 41].
More efforts have been made to the several kinds of AdS black
holes. The thermodynamic stability of the black holes constructed
in the general relativity and Gauss-Bonnet gravity respectively
was considered [63]. Both in the canonical ensembles and the grand
canonical ensembles, the analysis of black holes' thermodynamical
properties in de Rham, Gabadadze and Tolley (dRGT) massive gravity
was performed [64]. Chaturvedi et.al. considered the thermodynamic
geometry including the phase structure and critical phenomena for
four-dimensional Reissner-Nordstrom-AdS and Kerr-AdS black holes
in the canonical ensemble [65]. In the two kinds of ensembles, the
author of Ref. [67] derived and computed the thermodynamic
quantities of phantom AdS black holes to show the Hawking-Page
phase transition and the sign of heat capacity graphically [67].

It is necessary to scrutinize the thermodynamic characteristics of
Einstein-Euler-Heisenberg-AdS (EEH-AdS) black hole under different
conditions of fixed charge or fixed potential. The
self-interaction-corrected black holes need to be probed with
several respects. In addition to the motion of the particles with
the description of geodesics specified by the corrected black
holes, it is useful to study the thermodynamic quantities and
stability of the compact sources. We follow the procedure of Ref.
[40, 41] research on the EEH-AdS black holes with considerable
corrections from electromagnetic self-interactions. We wonder how
the electromagnetic self-interaction exerts the nonlinear
influence on the black holes just brings about some shifts on the
quantities or change the thermodynamic features of system greatly.
We derive the Hawking temperature, Helmholtz free energy, Gibbs
potential and entropy and then plot the dependence of the
quantities on the temperature under the nonlinear influence to
display the quantity difference evidently. The phase transition
appear and the black hole's stability can be confirmed. The
discussion and conclusion are listed in the end.

\section{The
	thermodynamic quantities of Einstein-Euler-Heisenberg-AdS black
	holes}

Within the frame of Euler-Heisenberg theory, the Lagrangian with
Lorentz and gauge invariants was constructed as [24, 25],
\begin{align}
\mathcal{L}_{EH}=-\frac{1}{4}F_{\mu\nu}F^{\mu\nu}
+\frac{\mu}{4}[(F_{\mu\nu}F^{\mu\nu})^{2}
+\frac{7}{4}(-^{\ast}F^{\mu\nu}F_{\mu\nu})^{2}]
\end{align}
where $\mu$ is the parameter demonstrating the
electromagnetic self-interactions,
\begin{align}
	\mu=\frac{2\alpha^{2}}{45m_{e}^{4}}
\end{align}
and $\alpha$ is the fine structure constant, $m_{e}$
being electron mass. It is noted that
$^{\ast}F^{\mu\nu}=\frac{1}{2\sqrt{-g}}\varepsilon_{\mu\nu\rho\sigma}
F^{\rho\sigma}$ is dual of $F_{\mu\nu}$, the electromagnetic field
tensor [24, 25]. The $\mu$-term in the Lagrangian (1) involves the
square of electromagnetic field tensor, leading the appearance of
nonlinear term for gauge field in the field equation [24, 25].

The four-dimensional action of general relativity coupled to the
nonlinear electromagnetic can be chosen as [24-26],
\begin{align}
	S_{EH}=\frac{1}{4\pi}\int_{M^{4}}d^{4}x\sqrt{-g}
	(\frac{R}{4}+\Lambda+\mathcal{L}_{EH})
\end{align}
resulting in the field equations, the
Einstein-Euler-Heisenberg equations [26, 30]. Here $g$ is the
determinant of the metric tensor and $R$ is the Ricci scalar. As a
solution to the field equations, the metric in a static
spherically symmetric form for Euler-Heisenberg-AdS black hole
surrounded by the strong electromagnetic field is [26, 28, 29],
\begin{align}
	ds^{2}=f(r)dt^{2}-\frac{dr^{2}}{f(r)}-r^{2}d\theta^{2}
	-r^{2}\sin^{2}\theta d\varphi^{2}
\end{align}
with the metric function,
\begin{align}
	f(r)=1-\frac{2M}{r}+\frac{Q^{2}}{r^{2}}-\frac{\mu Q^{4}}{20r^{6}}
	+\frac{r^{2}}{l^{2}}
\end{align}
where $M$ is the mass of the black hole and $Q$ is its
electric charge. Here $l$ is the AdS radius corresponding to the
cosmological constant like $\Lambda=-\frac{3}{l^{2}}$. It is noted
that $\mu$ is the Euler-Heisenberg parameter [26, 28, 29]. With
$\mu=0$, the metric (4) with its component (5) reduces to be
Reissner-Nordstrom ones [26, 28, 29]. Here the nonlinear effect
mainly appeared as $\mu$-term with the fourth power of electric
charge and the negative sixth power of the radial coordinate in
the spacetime metric is due to the vacuum polarization, leading
the compact object more gravitationally attractive than the
Reissner-Nordstrom-AdS black hole [26, 28, 29]. According to the
figures showing the behaviors of the metric functions of the
Einstein-Euler-Heisenberg black hole and Reissner-Nordstrom black
hole from Ref. [30], the shapes of the part of curves beyond the
outer horizons belonging to the two kinds of black holes
respectively are similar. Maybe the motion of test particles or
lights around the gravitational sources can not exhibit the subtle
difference between the black holes with or without the additional
corrections owing to the electromagnetic self-interactions
clearly. It is probable to compare the thermodynamic properties of
the Einstein-Euler-Heisenberg-AdS black holes with those of the
corresponding celestial bodies without nonlinear effect. We also
wonder whether the own features of the internal constitutions of
the nonlinear effect-corrected black holes may bring about their
evolution. It is indispensable to further the research on the
black holes under the nonlinear electrodynamics in the
thermodynamics direction. The horizon radii can be thought as real
roots of the equation like $f(r)=0$ [1]. Here we can let the
external horizon $r_{+}$ satisfy the condition as follows [30],
\begin{align}
	f(r_{+})=0
\end{align}
By combining the Eq. (5) and (6), the black hole mass
can be denoted as [30, 31],
\begin{align}
	M=\frac{r_{+}}{2}(1+\frac{Q^{2}}{r_{+}^{2}}-\frac{\mu
		Q^{4}}{20r_{+}^{6}}+\frac{r_{+}^{2}}{l^{2}})
\end{align}
According to the definition [31], the Hawking
temperature for this kind of black holes,
\begin{align}
	T=\frac{f'(r)}{4\pi}|_{r=r_{+}}\hspace{3cm}\notag\\
	=\frac{1}{4\pi r_{+}}(1-\frac{Q^{2}}{r_{+}^{2}} +\frac{\mu
		Q^{4}}{4r_{+}^{6}}+\frac{3r_{+}^{2}}{l^{2}})
\end{align}
while the entropy of black hole introduces [68],
\begin{align}
	S=\pi r_{+}^{2}
\end{align}
also written as $S=\frac{A}{4}$ where $A=4\pi r_{+}^{2}$
is the area of horizon [68]. The Hawking temperature versus the
horizon is depicted in the Figure 1. The asymptotic behavior of
the temperature are $\lim_{r_{+}\longrightarrow 0}T=\frac{1}{4\pi
	r_{+}}\frac{\mu Q^{4}}{4r_{+}^{6}}>0$ and
$\lim_{r_{+}\longrightarrow\infty}T=\frac{1}{4\pi
	r_{+}}\frac{3r_{+}^{2}}{l^{2}}>0$ which corresponding to the
curves in the Figure 1. The Figure 1 also demonstrates that the
curves with larger factor $\mu$ locate above the ones under the
weaker nonlinear effect in the small horizon scope and the curves
of Hawking temperature with different values of $\mu$ seem to
coincide in the case of huge black holes. Within the region with
small $r_{+}$, the temperature keeps positive, which is different
from Reissner-Nordstrom metric [40, 41]. It is manifest that the
curves of temperature function $T(r_{+})$ is concave in view of
the Figure 1. The condition equations are [40, 41],
\begin{align}
	\frac{\partial T}{\partial r_{+}}=0
\end{align}
\begin{align}
	\frac{\partial^{2}T}{\partial r_{+}^{2}}=0
\end{align}
Substituting the temperature (8) into the Eq.(10) and
Eq.(11), we obtain the minimum of Hawking temperature , charge and
horizon radius at the extreme point of temperature as follows,
\begin{align}
	r_{c}=\frac{1}{2}\frac{l^{2}+l\sqrt{l^{2}-7\mu}}
	{\sqrt{3(l^{2}+l\sqrt{l^{2}-7\mu})-7\mu}}
\end{align}
\begin{align}
	Q_{c}=\frac{1}{2\sqrt{2}}\frac{(l^{2}+l\sqrt{l^{2}-7\mu})^{\frac{3}{2}}}
	{3(l^{2}+l\sqrt{l^{2}-7\mu})-7\mu}
\end{align}
\begin{align}
	T_{c}=\frac{1}{28\pi l}\frac{32l+24\sqrt{l^{2}-7\mu}}
	{\sqrt{3(l^{2}+l\sqrt{l^{2}-7\mu})-7\mu}}
\end{align}
For the sake of the real quantities, the condition
$\mu\leq\frac{l^{2}}{7}$ must be obeyed. It is obvious that the
results from Eq.(7), Eq.(8), Eq.(12), Eq.(13) and Eq.(14) will
recover to be those in Ref.[40, 41, 67] when $\mu=0$. The equation
$T(r_{+})=T_{0}>T_{c}$ has two roots in contrast to the Figure 1.
\begin{figure}
	\centering
	\includegraphics[width=8cm]{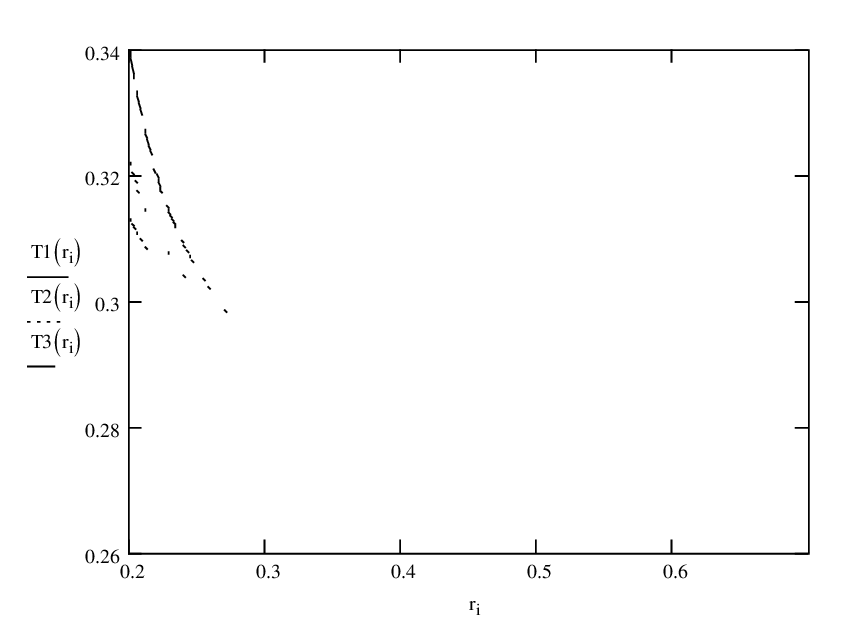}
	\caption{\label{fig:diagram} The solid, dotted and dashed curves corresponding to the
		dependence of Hawking temperature of Einstein-Euler-Heisenberg-AdS
		black holes in the canonical ensemble on the horizon for nonlinear
		factors $\mu=0.08, 0.1, 0.14$ respectively.}
\end{figure}

\section{The
	thermodynamic properties of Einstein-Euler-Heisenberg-AdS black
	holes in the canonical ensembles}

Now we have to investigate the specific thermal characteristics of
Einstein-Euler-Heisenberg-AdS black holes. It should be pointed
out that the gravitational sources involve the charges and the
sources evolve under different conditions [24-26, 28, 29]. We
start to discuss the thermodynamic stability and phase structure
of the black holes in the canonical ensemble where the charge is
invariant. According to the approach of Ref.[40, 41] and with the
help of Eq.(7), Eq.(8) and Eq.(9), the Helmholtz free energy
$F=M-TS$ is expressed as,
\begin{align}
	F=\frac{r_{+}}{4}(1+\frac{3Q^{2}}{r_{+}^{2}}-\frac{7\mu
		Q^{4}}{20r_{+}^{6}}-\frac{r_{+}^{2}}{l^{2}})
\end{align}
We connect the free energy (15) with the temperature (8)
to plot the curves with the self-interaction coupling $\mu$ in the
Figure 2. The shapes of curves with different values of parameter
$\mu$ are similar. There exists a point on the each curve as the
dependence of Helmholtz free energy on the Hawking temperature and
the free energy as a function of temperature is not derivable at
the point. The Helmholtz free energy is a multiple valued function
of Hawking temperature according to the Figure 2. When the
temperature is higher than the ones at the underivable point, a
series of curves for small free energy due to several valued $\mu$
are nearly identical. The underivable point of the Helnholtz free
energy can be thought as the critical point and certainly the
relevant temperature is the critical temperature denoted as
$T_{c}$.

\begin{figure}
	\centering
	\includegraphics[width=8cm]{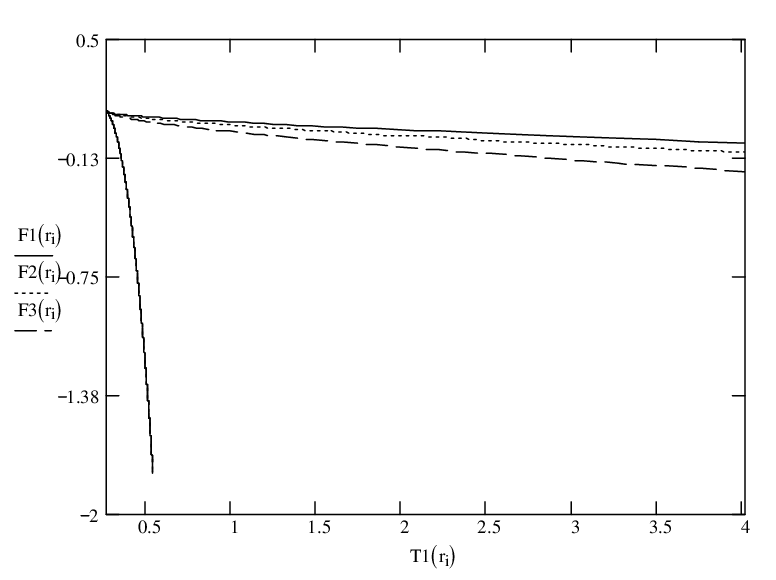}
	\caption{\label{fig:diagram} The solid, dotted and dashed curves of the Helmholtz free
		energy of Einstein-Euler-Heisenberg-AdS black holes in the
		canonical ensemble as functions of the temperature for nonlinear
		factors $\mu=0.08, 0.1, 0.14$ respectively.}
\end{figure}

It is noted that the black hole can satisfy the first law of
thermodynamics [69],
\begin{align}
	dM=TdS+\Phi dQ
\end{align}
where
\begin{align}
	\Phi=\frac{Q}{r_{+}}
\end{align}
is the electric potential. The corollary for the free
energy can be written as [69],
\begin{align}
	dF=-SdT+\Phi dQ
\end{align}

In the ensemble with fixed charge, the entropy of the
black hole is shown as [69],
\begin{align}
	S_{Q}=-(\frac{\partial F}{\partial T})_{Q}
\end{align}
In view of the temperature (8) and free energy (15), the
entropy of black hole like Eq.(9) can be varied again [68].
Because of the Eq.(8) and Eq.(19), the relation between the
entropy and temperature of the charged black holes within the
frame of nonlinear electrodynamics can be depicted in the Figure
3, Figure 4, Figure 5 with $\mu=0.08, 0.1, 0.14$ respectively. The
profiles of curves based on each diagram also resemble each other.
If the temperature approaches the critical value $T_{c}$, the
entropy (19) will become $\lim_{T\longrightarrow
	T_{c}}S_{Q}=-\frac{\lim_{T\longrightarrow T_{c}}(\frac{\partial
		F}{\partial r_{+}})_{Q}}{\lim_{T\longrightarrow
		T_{c}}(\frac{\partial T}{\partial
		r_{+}})_{Q}}\longrightarrow\infty$ because of the condition (10)
[40, 41]. It is discovered that the tangent line of the entropy
function at the critical point $T_{c}$ is perpendicular to the
$T$-axis in the figures. The critical point $T_{c}$ can be thought
as the common boundary. To each figure, we find that the entropy
curve starts to divide into two branches at the common boundary.
In addition the sufficiently high surface temperature of black
hole like $T>T_{c}$ corresponds to two values of the entropy.
Further the slopes of tangent lines for the upper branch keep
positive and the lower part reflects the decreasing function of
temperature with negative slopes. According to the black hole
entropy (9) from Ref.[68], the upper parts of entropies describe
the large black holes and the lower branch refers to the source
with small size.

Based on the analysis above, we continue analyzing the heat
capacity of the Einstein-Euler-Heisenberg black holes in the
ensemble. According to the thermodynamics of black holes [1, 31,
40, 41, 68], the black hole's heat capacity with fixed charge is
given by,
\begin{align}
	C_{Q}=T(\frac{\partial S_{Q}}{\partial T})_{Q}
\end{align}
Based on the three diagrams numbered 3, 4 and 5
respectively, it is clear that the temperature region formed with
$T\geq T_{c}$ means that the temperature keeps positive. The
slopes of tangent lines of the upper entropy curves are positive
while the slopes of lower part is negative. We can argue that the
large Einstein-Euler-Heisenberg-AdS black holes with positive heat
capacity are thermodynamically stable and the smaller ones with
negative one will evaporate instead of existing.

The stability of a black hole needs to be investigated in a
systematic way. The possibility for the fragmentation of a black
hole claimed that the black hole entropy must increase owing to
the second law of thermodynamics during its evolution [70]. Now we
wonder whether the charged black holes under the influence from
the self-interaction can break into pieces. The original black
hole can be thought as the initial state and certainly the final
state may consist of two black holes subject to the conservation
of mass and charge in the process of fragmentation. Within the
fragmentation, the mass and charge of the homogeneous source will
be rewritten as [70],
\begin{align}
	M=\varepsilon M+(1-\varepsilon)M
\end{align}
and
\begin{align}
	Q=\varepsilon Q+(1-\varepsilon)Q
\end{align}
respectively. The ratio is limited as
$0\leq\varepsilon\leq 1$ [71]. After splitting, one
Einstein-Euler-Heisenberg-AdS black hole has its metric like,
\begin{align}
	ds^{2}=f_{1}(r)dt^{2}-\frac{dr^{2}}{f_{1}(r)}-r^{2}d\theta^{2}
	-r^{2}\sin^{2}\theta d\varphi^{2}
\end{align}
where
\begin{align}
	f_{1}(r)=1-\frac{2\varepsilon M}{r}+\frac{\varepsilon^{2}Q^{2}}
	{r^{2}}-\frac{\mu\varepsilon^{4}Q^{4}}{20r^{6}}+\frac{r^{2}}{l^{2}}
\end{align}
with mass $\varepsilon M$ and charge $\varepsilon Q$.
The metric function $f_{2}(r)$ replaces $f_{1}(r)$ in the metric
(23) and the metric of the other part possessing mass
$(1-\varepsilon)M$ and charge $(1-\varepsilon)Q$ is obtained,
\begin{align}
	f_{2}(r)=1-\frac{2(1-\varepsilon)M}{r}+\frac{(1-\varepsilon)^{2}Q^{2}}
	{r^{2}}-\frac{\mu(1-\varepsilon)^{4}Q^{4}}{20r^{6}}+\frac{r^{2}}{l^{2}}
\end{align}
The metric with component function $f_{2}(r)$ belongs to
the background of the new Einstein-Euler-Heisenberg AdS black hole
that has mass $(1-\varepsilon)M$ and charge $(1-\varepsilon)Q$.
The entropy difference between the initial black hole and the two
separated parts can be expressed as [70],
\begin{align}
	\Delta S=\Delta S(\varepsilon)\hspace{1cm}\notag\\
	=(\pi r_{1}^{2}+\pi r_{2}^{2})-\pi r_{+}^{2}
\end{align}
The fragmented black holes have their own horizons
$r_{1}$ and $r_{2}$ satisfying $f_{1}(r_{1})=0$ and
$f_{2}(r_{2})=0$ respectively. The entropy difference between the
initial Einstein-Euler-Heisenberg-AdS black hole and the final
system including two black holes is shown graphically in the
Figure 6. The figure 6 declares that the difference keeps negative
no matter how considerable the nonlinear effect is. In the
ensemble, the division of this kind of AdS black holes can not
happen spontaneously. The nonlinear effect factor just adjust the
magnitude of the entropy difference within its region like
$\varepsilon\in[0, 1]$, but the sign of the difference remains
minus, so the charged AdS black holes involving electromagnetic
self-interactions will not split according to the second law of
thermodynamics. We can argue that the self-interactions may make
the curves for the dependence of entropy difference on the ratio
$\varepsilon$ different a little, but the nonlinear influence can
not change the curves greatly. Within the region of ratio
$\varepsilon$, the differences keep negative.

\begin{figure}
	\centering
	\includegraphics[width=8cm]{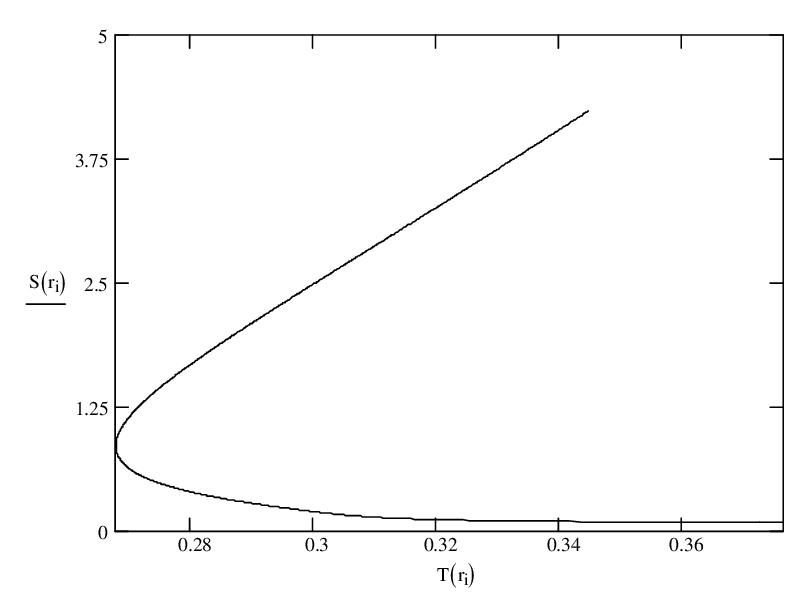}
	\caption{\label{fig:diagram} The curves of the entropy of
		Einstein-Euler-Heisenberg-AdS black holes in the canonical
		ensemble as functions of the temperature for nonlinear factors
		$\mu=0.08$.}
\end{figure}

\begin{figure}
	\centering
	\includegraphics[width=8cm]{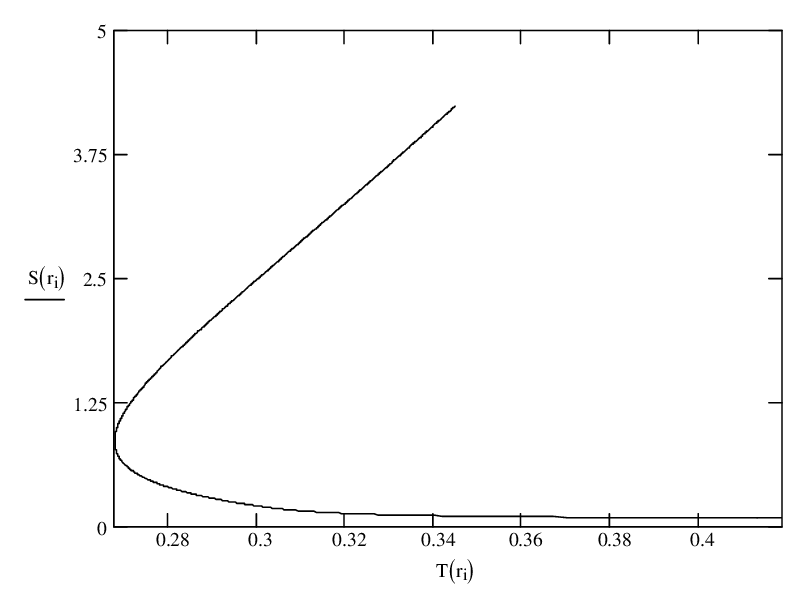}
	\caption{\label{fig:diagram} The curves of the entropy of
		Einstein-Euler-Heisenberg-AdS black holes in the canonical
		ensemble as functions of the temperature for nonlinear factors
		$\mu=0.1$.}
\end{figure}

\begin{figure}
	\centering
	\includegraphics[width=8cm]{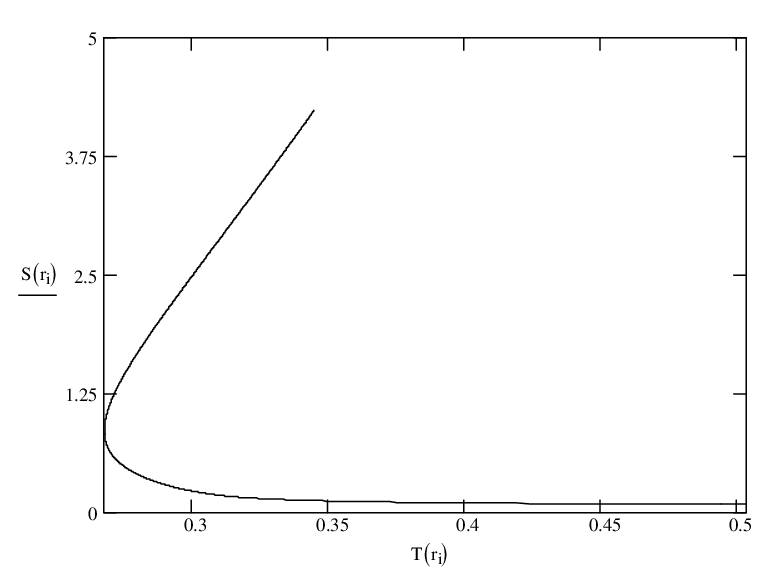}
	\caption{\label{fig:diagram} The curves of the entropy of
		Einstein-Euler-Heisenberg-AdS black holes in the canonical
		ensemble as functions of the temperature for nonlinear factors
		$\mu=0.14$.}
\end{figure}

\begin{figure}
	\centering
	\includegraphics[width=8cm]{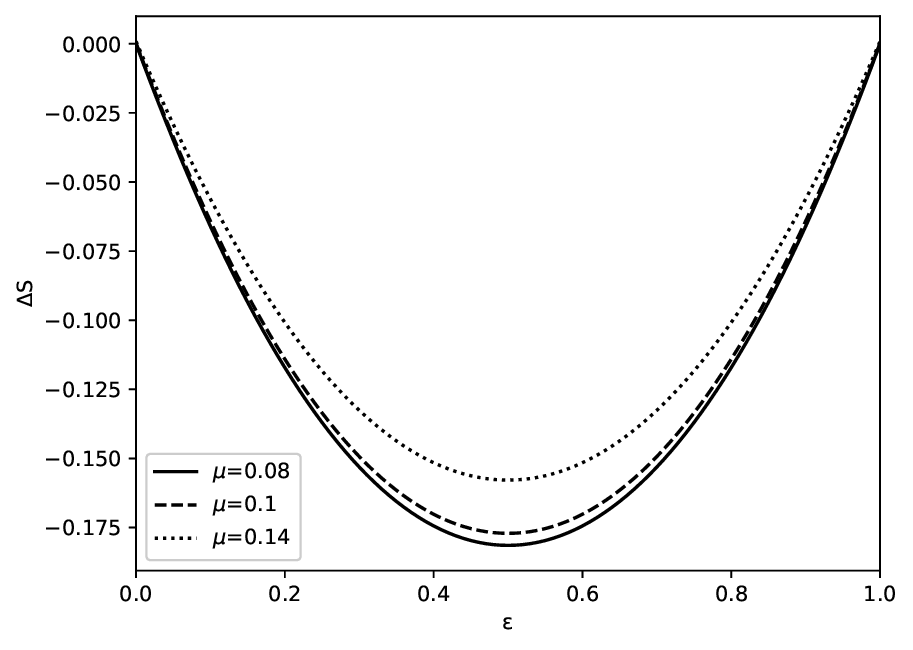}
	\caption{\label{fig:diagram} The solid, dashed and dotted curves of entropy difference
		between the initial Einstein-Euler-Heisenberg-AdS black holes and
		the fragmented bodies in the canonical ensemble as functions of
		the ration $\varepsilon$ for nonlinear factors $\mu=0.08, 0.1,
		0.14$ respectively.}
\end{figure}

\section{The
	thermodynamics properties of Einstein-Euler-Heisenberg-AdS black
	holes in the grand canonical ensembles}
	
It is necessary to discuss the stability and phase structure of
the charged black holes revised by the self-interactions when the
potential is invariant. By substituting the definition of electric
potential like Eq.(17) into the black hole's mass (17) and the
temperature (8), the corresponding thermodynamic variables like
the black hole's mass and its Hawking temperature are,
\begin{align}
	M=\frac{r_{+}}{2}(1+\Phi^{2}-\frac{\mu \Phi^{4}}{20r_{+}^{2}}
	+\frac{r_{+}^{2}}{l^{2}})
\end{align}
and
\begin{align}
	T=\frac{1}{4\pi r_{+}}(1-\Phi^{2}+\frac{\mu \Phi^{4}}{4r_{+}^{2}}
	+\frac{3r_{+}^{2}}{l^{2}})
\end{align}

In the process that the potential $\Phi$ is chosen to be
a value, the asymptotic behavior of the temperature is
$\lim_{r_{+}\longrightarrow 0}T=\frac{\mu\Phi^{4}}{16\pi
	r_{+}^{3}}>0$ for larger asymptotic value under stronger $\mu$
influence and $\lim_{r_{+}\longrightarrow\infty}T=\frac{3}{4\pi
	l^{2}r_{+}}>0$ which is equal to the asymptotic ones in the
canonical ensemble as discussed above. The Hawking temperatures
depending on the black hole horizon under the influence of
electromagnetic self-interactions are depicted in the Figure 7. It
should be pointed out that the curves shapes involving some
properties in the Figure 7 resemble those in the Figure 1. The
temperature curves are also concave, which is similar to the works
of the black holes in the fixed-charge ensembles. With the further
investigations on the Hawking temperature relating to the horizon,
the minimum temperature at the point is listed as follows,
\begin{align}
	T_{min}=T|_{r_{+}=r_{0}}\hspace{2cm}\nonumber\\
	=\frac{1}{4\pi r_{0}}(1-\Phi^{2}+\frac{\mu\Phi^{4}}{4r_{0}^{2}}
	+\frac{3}{l^{2}}r_{0}^{2})
\end{align}
with
\begin{align}
	r_{0}=\frac{l}{\sqrt{6}}\sqrt{1-\Phi^{2}+\sqrt{(1-\Phi)^{2}
			+\frac{9\mu\Phi^{4}}{l^{2}}}}
\end{align}

There are two valued $r_{+}$ to support the equation
$T(r_{+})>T_{min}$, one $r_{+}$ for smaller black hole and the
other for the larger one.

In the grand canonical ensembles, the Gibbs free energy is defined
as [40, 41],
\begin{align}
	G=M-TS-\Phi Q
\end{align}
From the first law of thermodynamics of black holes in
Eq.(16), the Gibbs potential function can be demonstrated as [40,
41],
\begin{align}
	dG=-SdT-Qd\Phi
\end{align}

In virtue of the mass (22) and temperature (23), the
Gibbs free energy is obtained,
\begin{align}
	G=\frac{r_{+}}{4}(1-\Phi^{2}-\frac{7\mu\Phi^{4}}{20r_{+}^{2}}
	-\frac{r_{+}^{2}}{l^{2}})
\end{align}

We show the Gibbs potential of the
nonlinear-effect-corrected black holes involving charges as
function of temperature graphically with a fixed electric
potential. Having investigated the curves with allowed values of
nonlinear factor $\mu$, we find that the shapes of the Gibbs
potential curves for different values of $\mu$ resemble each
other. The curve of Gibbs function with $\mu=0.1$ is shown in the
Figure 8. It should be stressed that the point of Gibbs free
energy at the minimum temperature is also underivable. The Gibbs
potential has also the unsmoothed connection, which is similar to
the Helmholtz free energy of the models in the canonical ensembles
as discussed above.

\begin{figure}
	\centering
	\includegraphics[width=8cm]{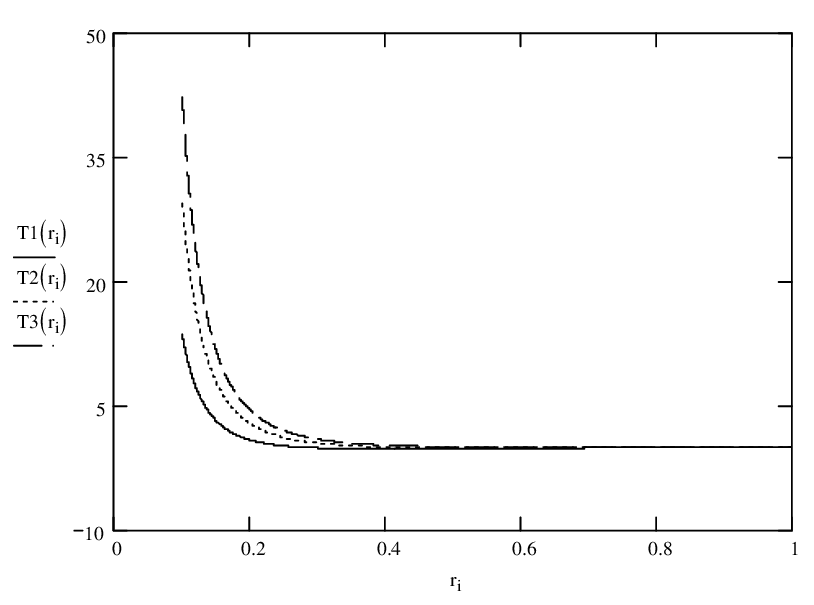}
	\caption{\label{fig:diagram} The solid, dotted and dashed curves corresponding to the
		dependence of Hawking temperature of Einstein-Euler-Heisenberg-AdS
		black holes in the grand canonical ensemble on the horizon for
		nonlinear factors $\mu=0.08, 0.1, 0.14$ respectively.}
\end{figure}

\begin{figure}
	\centering
	\includegraphics[width=8cm]{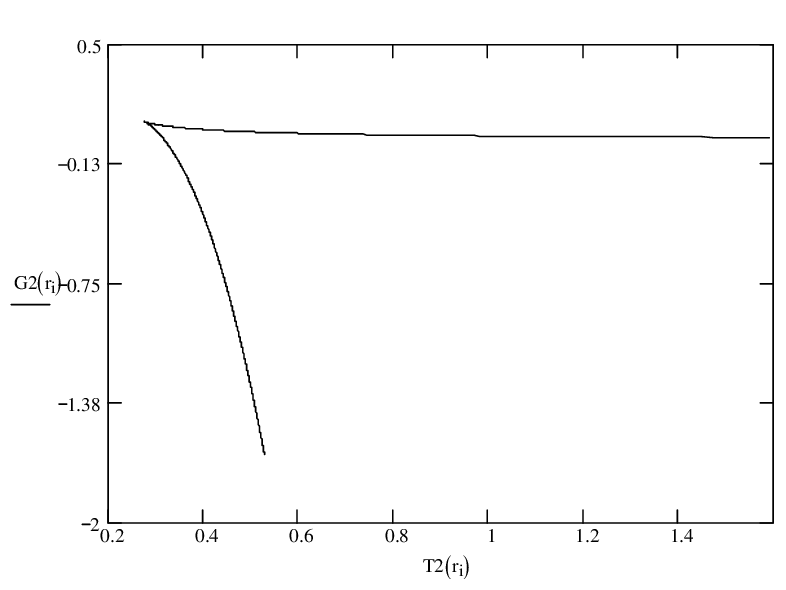}
	\caption{\label{fig:diagram} The curves of the Gibbs potential of
		Einstein-Euler-Heisenberg-AdS black holes in the grand canonical
		ensemble as functions of the temperature for nonlinear factors
		$\mu=0.1$.}
\end{figure}

\begin{figure}
	\centering
	\includegraphics[width=8cm]{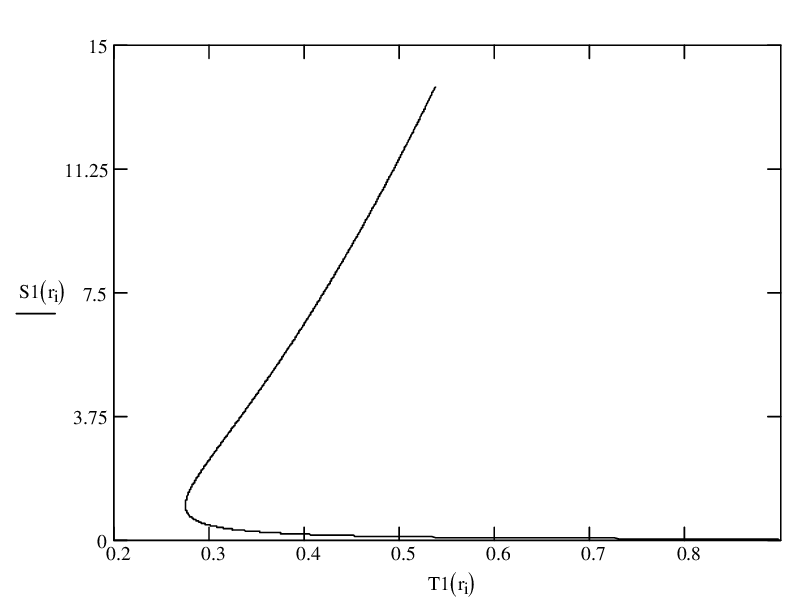}
	\caption{\label{fig:diagram} The curves of the entropy of
		Einstein-Euler-Heisenberg-AdS black holes in the grand canonical
		ensemble as functions of the temperature for nonlinear factors
		$\mu=0.08$.}
\end{figure}

\begin{figure}
	\centering
	\includegraphics[width=8cm]{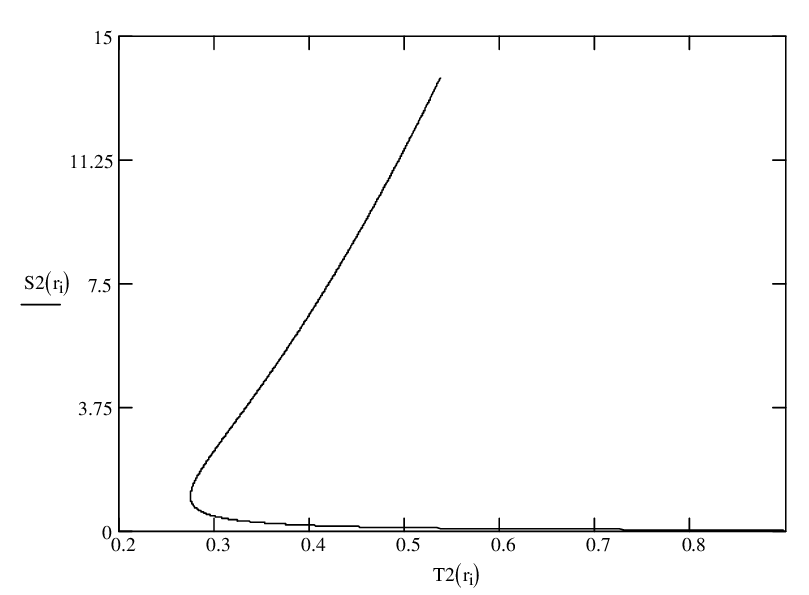}
	\caption{\label{fig:diagram} The curves of the entropy of
		Einstein-Euler-Heisenberg-AdS black holes in the grand canonical
		ensemble as functions of the temperature for nonlinear factors
		$\mu=0.1$.}
\end{figure}

\begin{figure}
	\centering
	\includegraphics[width=8cm]{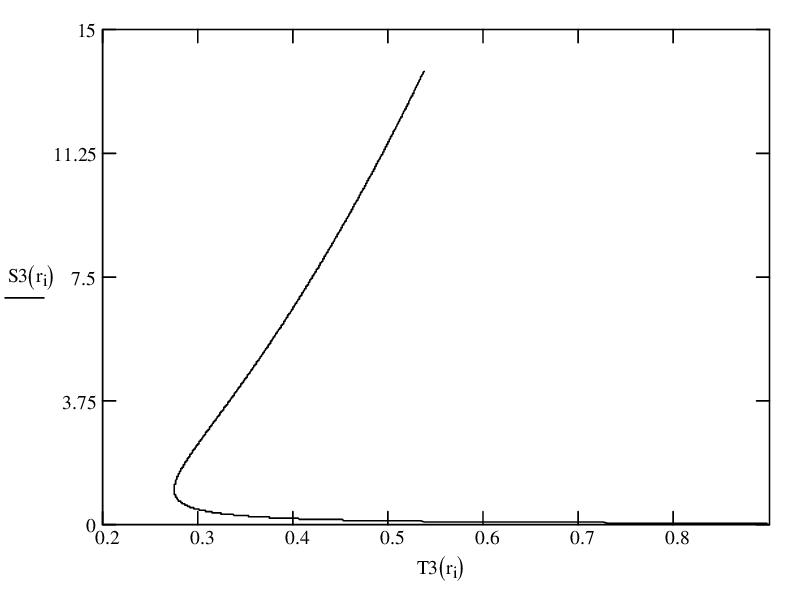}
	\caption{\label{fig:diagram} The curves of the entropy of
		Einstein-Euler-Heisenberg-AdS black holes in the grand canonical
		ensemble as functions of the temperature for nonlinear factors
		$\mu=0.14$.}
\end{figure}

We make use of the differential expression (32) to get the entropy
of the nonlinear-corrected and charged black hole [40, 41],
\begin{align}
	S_{\Phi}=-(\frac{\partial G}{\partial T})_{\Phi}
\end{align}

By means of the temperature (28) and the Gibbs potential
function (33), we plot the entropy function with various values of
$\mu$ specifying the nonlinear effects in the three diagrams
numbered as 9, 10, 11. A series of entropy curves are similar
although the parameter of self-interactions $\mu$ has different
values. It is interesting that the upper portion of the entropy
stands for the large black hole and the lower one belongs to the
compact object with small size. The phase transition will happen
at the joint of the two parts of entropy curves relating to the
temperature. The critical temperature $T_{c}$ for the underivable
point of Gibbs potential has something to do with the nonlinear
effect factor $\mu$. It is shown that the slopes of tangent lines
for the upper curves of entropy keep positive, meaning that the
relatively large charged black hole under nonlinear influence is
stable. The smaller black holes with negative capacity will
evaporate.

\section{Discussion and conclusion}

The thermodynamic quanties such as Hawking temperature, Helmholtz
free energy, Gibbs potential, entropy and heat capacity of the
Einstein-Euler-Heisenberg-AdS black holes are discussed in the
canonical ensembles and grand canonical ensembles respectively.
The black hole surface temperatures have the smallest value
associated with the horizon. The self-interactions of the
electromagnetic fields have an influence not to be ignored on the
smallest temperature. The phase transition may happen at the
lowest point. When the temperature is more than the smallest one,
the heat capacities of the relatively large black holes are
positive and the smaller black holes have negative heat
capacities. The tiny black holes with negative heat capacity
evaporate and will disappear in the evaporation. The huge ones
remain and the evolution proceeds. Most of black holes are
enveloped in the electromagnetic fields where the intensities have
something to do with the backgrounds like in the vicinity of the
centers of galaxies or neutron stars. The fields approaching the
critical strengths characterize the charged black holes [71]. The
charged black holes with self-interaction are universal and their
evolution and outcome need to be scrutinized. The expressions of
Hawking temperature at fixed-charge point or fixed-potential point
contain the parameter $\mu$ viewed as the self-interaction of
electromagnetic fields, so the self-interaction certainly brings
about the modifications to the thermodynamic quantities such as
Helmholtz free energy, Gibbs potential, entropy, heat capacity.
Our investigations show that the modified quantities due to
nonlinear effects result in the distinct corrections to the
thermodynamic characteristics and phase structure of the
Einstein-Euler-Heisenberg-AdS black holes in contrast to the
Reissner-Nordstrom-AdS black holes, but the influences from
nonlinear effect with allowed values of factor $\mu$ can not
change the evolution and the thermodynamic properties
fundamentally and thoroughly. This kind of huge black holes will
exist instead of evaporating and splitting.

\vspace{1cm}
\noindent \textbf{Acknowledge}

This work is partly supported by the Shanghai Key Laboratory of
Astrophysics 18DZ2271600.

\end{document}